\documentclass[12pt,preprint]{aastex}
\usepackage{graphicx}
\usepackage{natbib}

\shorttitle{GRAVITATIONAL WAVES FROM MERGING STRANGE STARS/PLANETS}
\shortauthors{Geng et al.}

\begin{document}

\title{Coalescence of Strange-Quark Planets with Strange Stars: \\
  a New Kind of Sources for Gravitational Wave Bursts}

\author{J. J. Geng\altaffilmark{1,2}, Y. F. Huang\altaffilmark{1,2}, and T. Lu\altaffilmark{3,4}}

\altaffiltext{1}{School of Astronomy and Space Science, Nanjing University, Nanjing 210046, China; hyf@nju.edu.cn}
\altaffiltext{2}{Key Laboratory of Modern Astronomy and Astrophysics (Nanjing University), Ministry of Education, Nanjing 210046, China}
\altaffiltext{3}{Purple Mountain Observatory, Chinese Academy of Sciences, Nanjing 210008, China}
\altaffiltext{4}{Joint Center for Particle, Nuclear Physics and Cosmology, Nanjing University - Purple Mountain Observatory, Nanjing 210093, China}

\begin{abstract}
Strange quark matter (SQM) may be the true ground state of hadronic matter, indicating that the
observed pulsars may actually be strange stars, but not neutron stars. According to this SQM
hypothesis, the existence of a hydrostatically stable sequence of strange quark matter stars has
been predicted, ranging from 1 --- 2 solar mass strange stars, to smaller strange dwarfs and even strange
planets. While gravitational wave (GW) astronomy is expected to open a new window to the universe, it will
shed light on the searching for SQM stars. Here we show that due to their extreme compactness, strange
planets can spiral very close to their host strange stars, without being tidally disrupted. Like
inspiraling neutron stars or black holes, these systems would serve as a new kind of sources for
GW bursts, producing strong gravitational waves at the final stage. The events occurring in
our local Universe can be detected by the upcoming gravitational wave detectors, such as
Advanced LIGO and the Einstein Telescope. This effect provides
a unique probe to SQM objects and is hopefully a powerful tool for testing the SQM hypothesis.
\end{abstract}


\maketitle

\section{INTRODUCTION}

With the operational and upcoming detectors, gravitational wave (GW) astronomy is expected to
open a new window to the universe in the near future. The last stage of inspiraling neutron
stars/black holes provides us a hopeful kind of candidates for GW sources \citep{Cutler93,Pozzo13}.
The Advanced LIGO \citep{Acernese06,Abbott09} detectors will be able to see inspiraling binaries
made up of two $1.4 M_{\odot}$ neutron stars to a distance of 300 Mpc.
This horizon distance would be promoted even to 3 Gpc by the future Einstein Telescope \citep{Punturo10,Hild08}.
In addition to these most promising candidates, people are eagerly looking for other potential GW
sources. For a normal matter planet moving around a compact star, the GW power is negligibly small
since the planet can not get very close to the central star as a whole due to the tidal disruption effect.
However, we argue that for a strange quark matter planet orbiting around a strange star (SS), the
corresponding GW signals can reach a detectable level. This is basically because the strange-matter
planet can get very close to the central compact star without being tidally disrupted, due to its
extreme compactness.

The existence of strange planets is based on a long-standing theory. It has long been proposed that
strange quark matter (SQM) may be the final ground state of hadronic matter \citep{Itoh70,Bodmer71,Witten84,Farhi84}.
Ordinary nuclei, made up of up and down quarks, may dissolve their boundaries and transit to a SQM phase
(consisted of up, down and strange quarks) if the nuclei are exerted to a high enough pressure.
Strange matter in bulk is stable. Even small chunks of strange matter with baryon number lower than $10^7$,
called ``strangelets'', may be stable due to the surface tension. If the SQM hypothesis is correct, then all
observed pulsars may actually be SSs but not neutron stars, due to the contamination process by strange
nuggets in the universe \citep{Alcock86}. Strange stars can exist in various forms, such as bare strange
stars or strange stars with a normal baryonic crust. Unlike neutron stars which have a critical
mass \citep{Chandra64}, there is no minimum mass for SSs.
Using the equation of state for SQM from phenomenological models, some authors have predicted the
existence of a hydrostatically stable sequence of SQM stars, ranging from strange dwarfs to strange
stars \citep{Glenden95a,Glenden95b,Vartanyan14}. It is interesting to note that strange planets exist
in this continuous sequence.

How to identify strange-matter objects or test the SQM hypothesis?
Currently, several possible ways have been proposed. According to the equation of state of SQM in the
MIT Bag model \citep{Farhi84,Krivoru91,Madsen99}, the mass -- radius relation for SSs
follows $M \propto R^3$ if $M < 1 M_{\odot}$, very different from $M \propto R^{-3}$ for neutron stars.
Unfortunately, for SSs and neutron stars with the same mass of $\sim 1.4 M_{\odot}$, their radii are
similar. Observations show that the average mass of pulsars is around $1.4 M_{\odot}$ \citep{Lattimer07},
consequently leading to the limitation of this method \citep{Panei00}. Later, it has been argued that the
high cooling rate of SQM together with quick thermal response of the thin crust yields low surface
temperatures of SSs as compared to neutron stars of the same age \citep{Pizzoch91}. However, the cooling
rate of neutron stars could also be high after considering more details \citep{Page92,Lattimer94},
reducing the temperature differences between the two kinds of objects. Noting the larger shear and bulk
viscosities in SSs, some researchers also suggested that they can spin more rapidly, more approaching
the Kepler limit \citep{Frieman89,Glenden89,Friedman89}. If the spin period of a young pulsar is less
than 1 ms, then it is very likely to be an SS rather than a neutron star \citep{Kristian89}. But these
fast spinning objects themselves are difficult to be detected observationally. Researchers also noticed
GWs as a possible tool for probing SSs. As rotating relativistic stars, SSs can emit GWs due to normal
mode or r-mode oscillations \citep{Madsen98,Andersson01,Lindblom00,Andersson02} or global solid
deformation \citep{Jaranow98,Jones02}. However, these GWs are generally very weak and the difference
between SSs and NSs are even smaller. Anyway, it is interesting to note that an upper limit of $10^{-24}$
for the GW strain amplitude have been obtained for 28 known pulsars \citep{Abbott05}. Also, GW signals
of SS mergers may differ slightly from those of neutron star mergers \citep{Bauswein10,Moraes14}, but
the difference is also difficult to measure. In short, despite long lasting and extensive investigations,
the task of identifying strange-matter objects or testing the SQM hypothesis still remains a challenge
for researchers hitherto.

In this work, we study the last stage of the inspiraling of a strange-matter planet toward a strange
star. Very different to what happens in the counterpart of a neutron star planetary system, the strange
planet can get very close to the host strange star without being tidally disrupted, forming a minitype
double compact star system. As a result, an eminent GW burst will be generated due to the final merge.
We show that GW emission from these events happening in our local Universe is strong enough to be
detected by the upcoming detectors such as Advanced LIGO and the Einstein Telescope. Such an effect
can be used as a unique probe to the existence of SQM stars.

\section{GWS FROM MERGING SQM STARS/PLANETS}

\subsection{Strain Amplitude Evolution}

Let us consider a binary system composed of two members with masses $M$ and $m$ respectively.
For simplicity, we assume that the primary compact star has a mass of $M = 1.4 M_{\odot}$ and the
companion star is a planet so that $m \ll M$. The GW radiation power from this binary system is then
\begin{equation}
P = \frac{32 G^4 M^2 m^2 (M+m)}{5 c^5 a^5},
\end{equation}
where $G$ is the gravitational constant, $c$ is the light velocity and $a$ is the semi-major axis.
The measurable signals of GWs are the amplitudes of two polarized components --- $h_{+}$ and $h_{\times}$.
For merging binaries, we assume the waves to be sinusoidal and define an effective
strain amplitude as $h = (\left<h^2_{+}\right> + \left<h^2_{\times}\right>)^{1/2}$. After averaging over
the orbital period, we can obtain \citep{Peters63,Press72,Postnov14}
\begin{equation}
h = 5.1 \times 10^{-23}  \left(\frac{\mathcal{M}}{1~M_{\odot}}\right)^{5/3}
  \left(\frac{P_{\rm orb}}{1~\rm hr}\right)^{-2/3}  \left(\frac{d}{10~\rm{kpc}}\right)^{-1},
\end{equation}
where $\mathcal{M} = (M m)^{3/5} / (M+m)^{1/5}$ is the chirp mass, $P_{\rm orb}$ is the orbital
period and $d$ is the distance of the binary to us.

If the planet is a normal-matter one, the GW signals will always be extremely weak because the planet
cannot come very close to the compact primary star. The strong tidal force from the central object
will disrupt the planet when it is still far away. Assuming a density of $\rho_0$, a normal planet
will be disrupted at the distance of
\begin{equation}
r_{\rm td} \approx 5.1 \times 10^{10} \left(\frac{M}{1.4~M_{\odot}}\right)^{1/3} \left(\frac{\rho_0}{10~\rm{g~cm}^{-3}}\right)^{-1/3} \rm{cm}.
\end{equation}
$r_{\rm td}$ is usually called the tidal disruption radius. If $r_{\rm td}$ is too large,
the GW emission will be very weak.
For example, for a normal planet of $m = 10^{-6} M_{\odot}$ (with density
$\rho_0 \sim 10~\rm{g~cm}^{-3}$ and radius $R \sim 3.6 \times 10^8$ cm) disrupted
at $ 5.1 \times 10^{10}$ cm, the maximum GW amplitude is only  $h \approx 4.9 \times 10^{-29}$ (with
a very low frequency of $3.8 \times 10^{-4}$ Hz) at a distance of 10 kpc, which is too weak to be
detected. Even for a giant normal-matter planet of $m = 10^{-3} M_{\odot}$, the maximum GW amplitude
is again only $h \approx 4.9 \times 10^{-26}$, which is still far beyond the detection limit.

However, when the companion is a strange-matter planet, things will become very different.
According to the canonical MIT Bag model for SQM, the mass -- radius relation of strange stars
can be well described by $m \propto R^3$. This relation applies to the whole sequence of
bare strange stars, including strange planets. The extreme high
density (typically $\rho_0 = 4.0 \times 10^{14}~\rm{g~cm}^{-3}$)
of strange planet ensures that it can come very close to the compact host star while retaining
its integrity, because the tidal disruption radius now becomes $r_{\rm td} = 1.5 \times 10^6$ cm.
 This will give birth to a minitype double compact star system, which will be very efficient in
 producing GWs. At the last stage of the inspiraling (i.e. when the planet approaches the tidal
disruption radius, $r_{\rm td}$), the strain amplitude of GWs from a strange-matter binary system is
\begin{eqnarray}
h &=& 1.4 \times 10^{-24} \left(\frac{M}{1.4~M_{\odot}}\right)^{2/3} \left(\frac{\rho_0}{4.0 \times 10^{14}~\rm{g~cm}^{-3}}\right)^{4/3} \nonumber\\
&& \times \left(\frac{R}{10^4~\rm{cm}}\right)^{3} \left(\frac{d}{10~\rm{kpc}}\right)^{-1}.
\end{eqnarray}
According to this equation, the strain amplitude of GWs from a strange planet of
mass $m = 10^{-4} M_{\odot}$ ($R = 5.0 \times 10^4$ cm, $\rho_0 = 4.0
\times 10^{14}~\rm{g~cm}^{-3}$) will be $1.7 \times 10^{-22}$ at a distance of $\sim 10$ kpc from
us. This amplitude is comparable to that of a neutron star -- neutron star binary system (when the
orbital period is around 1 s) at $\sim 1$ Mpc. So, such strange star -- strange planet systems would
be appealing targets for the ongoing and upcoming GW experiments, such as Advanced LIGO and the
Einstein Telescope.

\begin{figure}
\center
\includegraphics[width=0.7\linewidth]{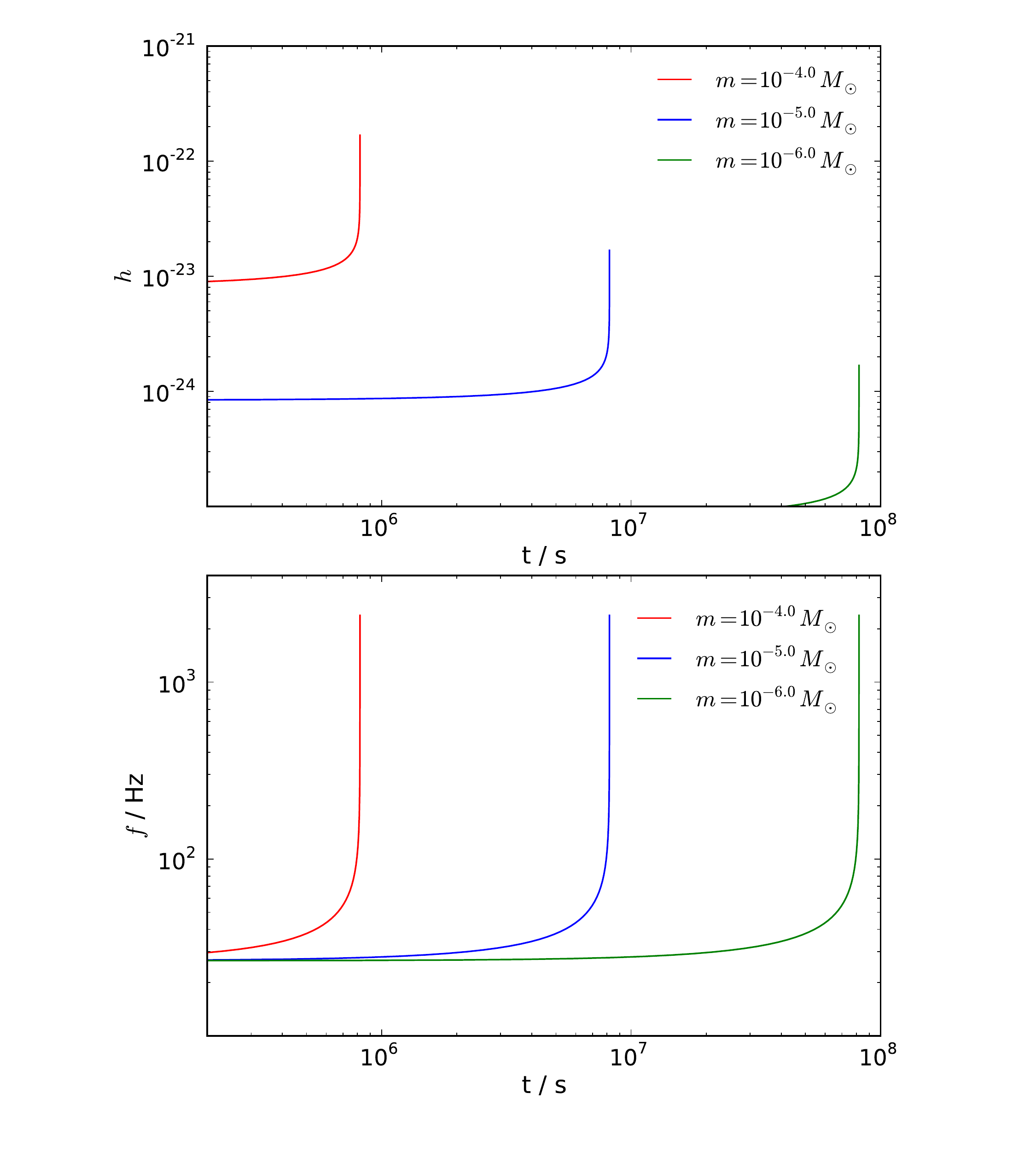}
\caption{\label{fig:1} Evolution of GW amplitude (upper panel) and frequency (lower panel) for coalescing
strange star - strange planet systems at a distance of 10 kpc from us. 
The host strange star has a mass of $1.4 M_{\odot}$. The orbits are assumed to be circular, and the
initial separation of the two objects is set to be $20 r_{\rm td}$ ($r_{\rm td}$ is the tidal disruption radius).
The solid red, blue, and green lines correspond to strange planets with a mass of $10^{-4} M_{\odot}$,
$10^{-5} M_{\odot}$, and $10^{-6} M_{\odot}$ respectively. For all the strange planets,
a mean density of $\rho_0 = 4.0 \times 10^{14} \rm{g}~\rm{cm}^{-3}$ has been taken. 
In all the cases, we stop our calculations at the tidal disruption radius, which 
gives the highest GW amplitude and also the highest GW frequence at the end point of each curve.
}

\end{figure}

Since the inspiraling is a gradual process during which the strange planet approaches the central
strange star progressively, we need to consider the evolution of the GW amplitude in the whole
procedure. Assuming that the orbit always keeps to be circular, the emission power of GWs can be
calculated according to Equation (1). We can then easily know how quickly the orbit shrinks and
how the GW amplitude evolves. In Fig.~\ref{fig:1}, the evolution of $h$ during the inspiraling is
illustrated (assuming a distance of 10 kpc from us), with three different masses assumed for
the strange planets. Correspondingly, the evolution of the GW frequency 
($f = 2 / P_{\rm orb}$) is also shown in the lower panel. 
In our calculations, in order to focus on the last stage of the coalescence we have 
taken $20 r_{\rm td}$ as the initial separation between the two objects. Note that the GW emission is 
usually very weak and also evolves very slowly when $r < 20 r_{\rm td}$. 
It can be clearly seen that in all these cases,
the GW signal can rise to a high level at the last stage of the coalescence process. For example, in the
$m = 10^{-6} M_{\odot}$ case, $h$ can remain to be larger than $10^{-24}$ for a long time of 3500 s.
The GW frequency of these systems is also in the most sensitive range of LIGO and Einstein Telescope, making
them very appealing GW sources.

\subsection{Strain Spectral Amplitude}

\begin{figure}
\includegraphics[width=\linewidth]{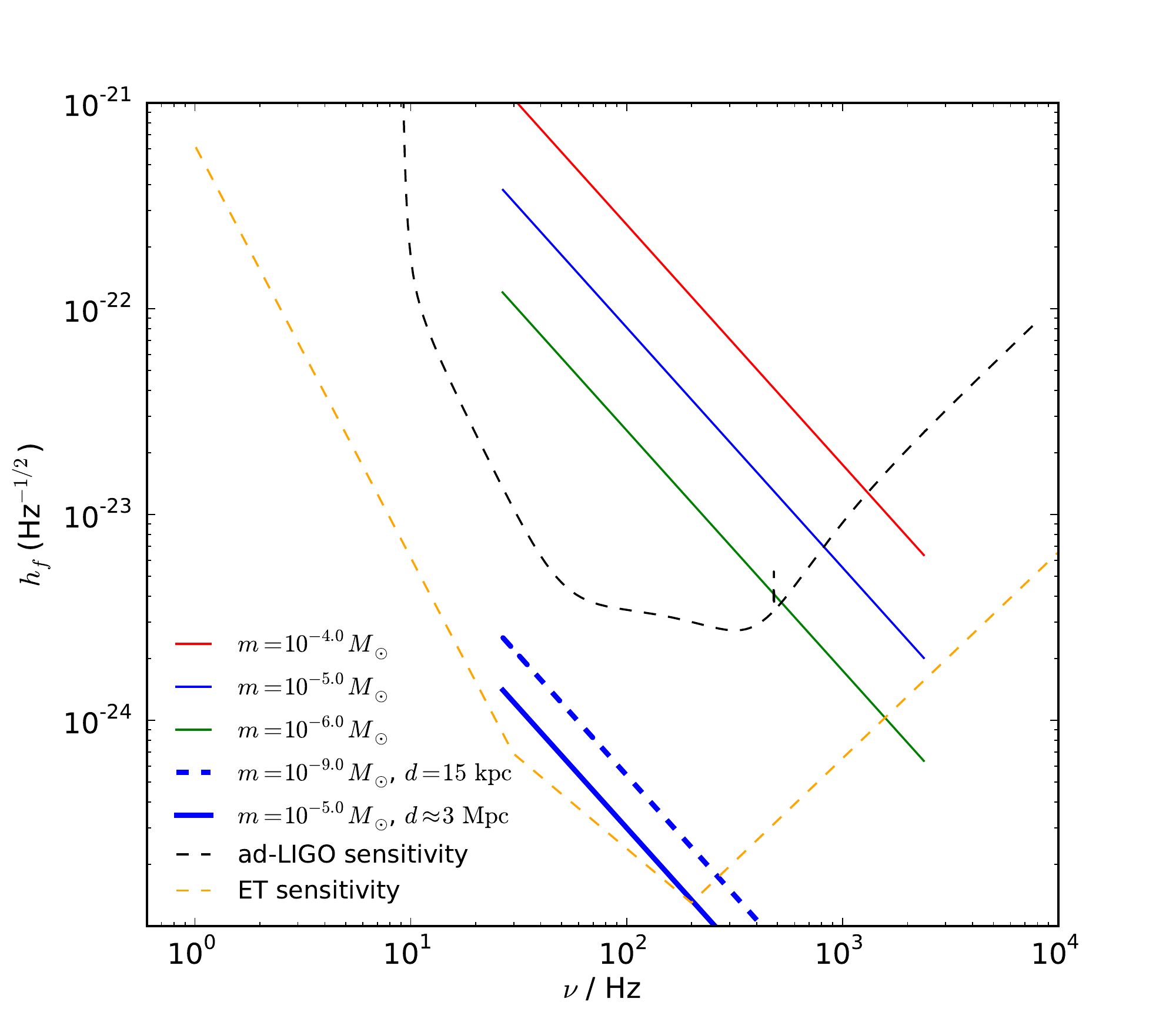}
\caption{\label{fig:2} Strain spectral amplitude of the GWs against frequency for coalescing strange
star - strange planet systems. The host strange star has a mass of $1.4 M_{\odot}$. The straight red,
blue, and green solid lines correspond to strange planets with a mass of $10^{-4} M_{\odot}$,
$10^{-5} M_{\odot}$, and $10^{-6} M_{\odot}$ respectively, with the system lying at a
distance of 10 kpc from us. 
In these cases, we stop our calculations at the tidal disruption radius, which 
gives the highest GW frequence at the end point of each curve.
The thick blue dashed line corresponds to a strange planet mass of $10^{-9} M_{\odot}$ and with 
the system at 15 kpc from us.
The thick blue solid line corresponds to a strange planet mass of $10^{-5} M_{\odot}$,
but with the system residing at a distance of 3 Mpc. The results are compared with the sensitivity curves
of Advanced LIGO (the dashed black curve, \citet{Harry10})
and future Einstein Telescope (the dashed orange curve, \citet{Hild08}).}
\end{figure}

To judge whether the GWs could be detected by GW experiments, it is useful to plot it against the
sensitivity curve. This is usually done by considering the strain spectral amplitude ($h_{f}$), which
is defined as the square root of the power spectral density, i.e. the power per unit frequency.
For the double compact star systems studied here, the Fourier transform of $h(t)$ can be found
when $f$ changes slowly \citep{Finn93,Nissanke10,Postnov14}, which is
\begin{eqnarray}
h_f &=& 6.4 \times 10^{-21} \left(\frac{\mathcal{M}}{1~M_{\odot}}\right)^{5/6} \left(\frac{f}{300~\rm{Hz}}\right)^{-7/6} \nonumber\\
&& \times \left(\frac{d}{10~\rm{kpc}}\right)^{-1} \rm{Hz}^{-1/2}.
\end{eqnarray}
Using this stationary phase approximation, the strain spectral amplitude against frequency is plotted in Fig.~\ref{fig:2},
together with the sensitivity curves of Advanced LIGO and the Einstein Telescope. It can be clearly seen that
the GW signals from the coalescing strange star systems in our Galaxy 
(with the planet mass larger than $\sim 10^{-9}$ M$_\odot$)
can be well detected by these experiments.
More encouragingly, the horizon distance of Einstein Telescope to these events (assuming $m \geq 10^{-5} M_{\odot}$)
will even be $\sim 3$ Mpc, which means the mergers happening in nearby galaxies will also be spotted.

\begin{figure}
\includegraphics[width=\linewidth]{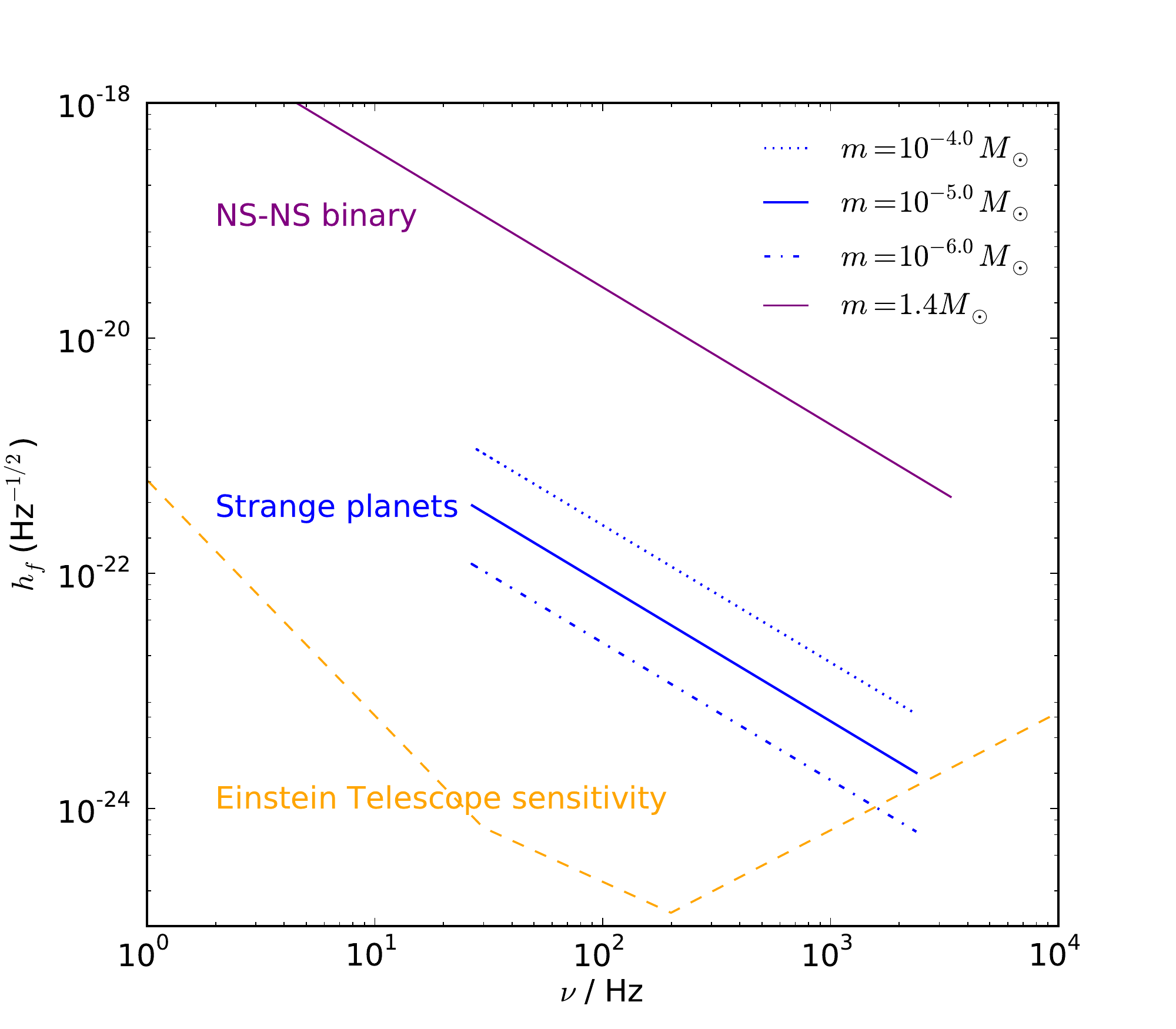}
\caption{\label{fig:3}Strain spectral amplitude of the GWs vs. frequency for different binary compact star systems.
For strange star - strange planet systems (blue lines), three different planet masses are assumed. The distance is
taken as 10 kpc from us. As a direct comparison, we also plot the case of a double neutron star system (the purple
line), again residing at 10 kpc. 
In all the cases, we stop our calculations at the tidal disruption radius, which 
gives the highest GW frequence at the end point of each curve.
The sensitivity curve of Einstein Telescope is shown by the dashed orange curve.
It can be seen that the GW emission from a double neutron star system (with two compact
stars that are both about 1.4 M$_\odot$) are much stronger than a strange star - strange planet system at the
same distance. Thus these two kinds of GW sources can be easily discriminated observationally.}
\end{figure}

It is interesting to note that high quality GW observations of binary compact star coalescences can directly
provide their distance information \citep{Schutz86,Messen12}, because both the GW amplitude and the frequency
evolution can be measured during the inspiraling process. On the other hand, the distance may also be
determined by electromagnetic observations on the counterparts, since the coalescence is likely to lead
to a strong hard X-ray burst \citep{Huang14}. In the future, if a GW signal of an appropriate amplitude
is detected from our local Universe, it would most likely come from the merge of a strange planet with its
host strange star (if also happened in our local Universe, the GWs from a double neutron star system
will be much stronger and easy to discriminate, see Fig.~\ref{fig:3}).
It then can be regarded as a strong proof for the existence of SQM.

\section{CONCLUSIONS AND DISCUSSION}

In this study, we have calculated the GW signals from strange star - strange planet systems during the
final inspiraling phase. The high density of the strange-matter planet ensures it to survive the
tidal disruption and come very close to the compact central star, leading to strong GW emission.
Our results indicate that strange planets with $m \geq 10^{-5} M_{\odot}$ can result
in GW outbursts detectable by the future Einstein Telescope up to a horizon of 3 Mpc.
These events comprise a completely new kind of GW sources, which, if detected, will be strong
evidence supporting the SQM hypothesis.

Our calculations are based on the assumption of the existence of strange star - strange planet
systems. There are at least three possible scenarios in which such systems may be generated.
First, newly-born strange-quark stars are likely to be hot and highly turbulent. They
may eject low-mass quark nuggets. It has been suggested that ejection of planetary clumps may happen simultaneously
during the formation of a strange star due to strong turbulence on the surface \citep{Xu03,Xu06,Horvath12}. If the
ejected strange quark planet is somehow gravitationally bounded, then a strange planetary system can be
directly formed. In this case, a convective velocity lager than $10^9$ cm s$^{-1}$ on the surface would be needed
for the ejection. Second, another possible scenario involves the contamination processes. During the supernova
explosion that gives birth to a strange star, if the planets of the progenitor star can survive the violent
process (do not escape or be vaporized), then they may be contaminated by the abundant strange quark nuggets
ejected from the newly-born strange star and be converted to strange planets. In fact, two planets of a few
Earth-mass have been confirmed orbiting around the pulsar PSR B1257+12 \citep{Wolszczan92}.
If these planets are remnants of the progenitor star, then the possibility that they have been contaminated
and converted to strange planets cannot be excluded currently \citep{Caldwell91,Glenden95a,Madsen99}.
Finally, according to the Big Bang theory, our Universe once experienced a so called quark phase stage, during which the
density and temperature were both extremely high. Planetary strange-matter objects may be directly formed at that
stage and may survive till now \citep{Cott94}. Such objects could be very numerous and make up the dark objects in
galactic halos \citep{Chandra00}. They can be captured by strange stars or neutron stars to form planetary systems.

It is interesting to know how many GW bursts from strange planetary systems could be observed by future 
GW telescopes each year. Following the idea that SQM be the final ground state of hadronic matter, 
we assume that all neutron stars are truely strange stars. It is estimated that there are about 
$10^9$ NSs in our Milky Way galaxy \citep{Timmes96}, so we take this number as the total amount of strange stars 
in our Galaxy. From a conservative view, planetary systems appear to occur in around $10^{-3}$ of 
pulsars \citep{Wolszczan92,Greaves00}. Thus the number of strange planetary systems in our Galaxy could 
be $\sim 10^9 \times 10^{-3} \sim 10^6$. On the other hand, it has been argued that planets around a 
compact star may collide with each other, generating some large fragments which can fall onto the 
central compact star. This mechanism has been suggested to account for the bursts from some soft gamma 
repeaters \citep{Katz94}. According to Katz et al. (1994), the generated solid bodies can be in the 
mass range of $10^{22}$ g --- $10^{25}$ g. In our study, in order for the GW bursts to be detectable 
within our Galaxy, the solid body should be larger than $\sim 10^{24}$ g (or $\sim 10^{-9}$ M$_\odot$, 
see Fig. 2). Following the derivations of Katz et al. (1994), the timescale for a single planetary 
system to undergo such a collision is then $\sim 10^5$ years. Finally, we estimate that 
$\sim 10^6/10^5 = 10$ coalescence events could be detected as GW bursts by future Einstein
Telescope. Additionally, Fig. 2 clearly shows that the horizon distance for some large strange 
planets can be as far as a few Mpc. Thus many coalescence events in our local Universe (from other 
nearby galaxies) will also be detectable. So, we believe that the derived event rate of 
about 10 per year is still the lower limit.

Finally, it is worthy to note that many low-mass black hole binary systems may exist in our Milky Way
and nearby galaxies \citep{Nakamura97}. These systems can also merge and give birth to GW bursts with the 
amplitudes similar to the events discussed here. Several methods may help us to discriminate these two 
kinds of GW sources. First, the chirp mass of a strange star - strange planet system is usually much 
smaller than that of typical low-mass black hole binaries. So the chirp mass measured from the GW signals
can help to distinguish them. Second, some forms of hard X-ray bursts may associate with the coalescence of 
strange star - strange planet systems and can be basically observed, while no significant electromagnetic 
emission is expected from the mergers of low-mass black holes. Finally, for the strange star - strange planet
system, the strange star can be observed as a pulsar before the coalescence. After the coalescence, the strange 
star is retained and it still can show up as a pulsar. However, for a low-mass black hole binary, usually no 
electromagnetic counterpart can be directly detected before and after the coalescence.

\begin{acknowledgments}
We are very grateful for the valuable suggestions from the anonymous referee. 
We thank Yiming Hu for helpful discussions. 
This work was supported by the National Basic Research Program of China with Grant No.
2014CB845800, and by the National Natural Science Foundation of China with Grant No. 11473012.
\end{acknowledgments}

\label{lastpage}

\begin{thebibliography}{}
\bibitem[Abbott et al.(2005)]{Abbott05} Abbott, B., Abbott, R.,
Adhikari, R., et al.\ 2005, Physical Review Letters, 94, 181103

\bibitem[Abbott et al.(2009)]{Abbott09} Abbott, B.~P., Abbott,
R., Acernese, F., et al.\ 2009, \nat, 460, 990

\bibitem[Acernese et al.(2006)]{Acernese06} Acernese, F., Amico,
P., Alshourbagy, M., et al.\ 2006, Classical and Quantum Gravity, 23, 635

\bibitem[Alcock et al.(1986)]{Alcock86} Alcock, C., Farhi, E.,
\& Olinto, A.\ 1986, \apj, 310, 261

\bibitem[Andersson et al.(2002)]{Andersson02} Andersson, N., Jones,
D.~I., \& Kokkotas, K.~D.\ 2002, \mnras, 337, 1224

\bibitem[Andersson
\& Kokkotas(2001)]{Andersson01} Andersson, N., \& Kokkotas, K.~D.\ 2001, International Journal of Modern Physics D, 10, 381

\bibitem[Bauswein et al.(2010)]{Bauswein10} Bauswein, A.,
Oechslin, R., \& Janka, H.-T.\ 2010, \prd, 81, 024012

\bibitem[Bodmer(1971)]{Bodmer71} Bodmer, A.~R.\ 1971, \prd, 4,
1601

\bibitem[Caldwell
\& Friedman(1991)]{Caldwell91} Caldwell, R.~R., \& Friedman, J.~L.\ 1991, Physics Letters B, 264, 143

\bibitem[Chandra
\& Goyal(2000)]{Chandra00} Chandra, D., \& Goyal, A.\ 2000, \prd, 62, 063505

\bibitem[Chandrasekhar(1964)]{Chandra64} Chandrasekhar, S.\ 1964,
Physical Review Letters, 12, 114

\bibitem[Cottingham et al.(1994)]{Cott94} Cottingham, W.~N.,
Kalafatis, D., \& Vinh Mau, R.\ 1994, Physical Review Letters, 73, 1328

\bibitem[Cutler et al.(1993)]{Cutler93} Cutler, C., Apostolatos,
T.~A., Bildsten, L., et al.\ 1993, Physical Review Letters, 70, 2984

\bibitem[Del Pozzo et al.(2013)]{Pozzo13} Del Pozzo, W., Li,
T.~G.~F., Agathos, M., Van Den Broeck, C.,
\& Vitale, S.\ 2013, Physical Review Letters, 111, 071101

\bibitem[Farhi
\& Jaffe(1984)]{Farhi84} Farhi, E., \& Jaffe, R.~L.\ 1984, \prd, 30, 2379

\bibitem[Finn
\& Chernoff(1993)]{Finn93} Finn, L.~S., \& Chernoff, D.~F.\ 1993, \prd, 47, 2198

\bibitem[Friedman et al.(1989)]{Friedman89} Friedman, J.~L.,
Ipser, J.~R., \& Parker, L.\ 1989, Physical Review Letters, 62, 3015

\bibitem[Frieman
\& Olinto(1989)]{Frieman89} Frieman, J.~A., \& Olinto, A.~V.\ 1989, \nat, 341, 633

\bibitem[Glendenning et al.(1995a)]{Glenden95a} Glendenning, N.~K.,
Kettner, C., \& Weber, F.\ 1995, Physical Review Letters, 74, 3519

\bibitem[Glendenning et al.(1995b)]{Glenden95b} Glendenning, N.~K.,
Kettner, C., \& Weber, F.\ 1995, \apj, 450, 253

\bibitem[Glendenning(1989)]{Glenden89} Glendenning, N.~K.\ 1989,
Physical Review Letters, 63, 2629

\bibitem[Greaves
\& Holland(2000)]{Greaves00} Greaves, J.~S., \& Holland, W.~S.\ 2000, \mnras, 316, L21 

\bibitem[Harry
\& LIGO Scientific Collaboration(2010)]{Harry10} Harry, G.~M., \& LIGO Scientific Collaboration 2010, Classical and Quantum Gravity, 27, 084006

\bibitem[Hild et al.(2008)]{Hild08} Hild, S., Chelkowski, S.,
\& Freise, A.\ 2008, arXiv:0810.0604

\bibitem[Horvath(2012)]{Horvath12} Horvath, J.~E.\ 2012, Research
in Astronomy and Astrophysics, 12, 813 

\bibitem[Huang
\& Geng(2014)]{Huang14} Huang, Y.~F., \& Geng, J.~J.\ 2014, \apjl, 782, L20

\bibitem[Itoh(1970)]{Itoh70} Itoh, N.\ 1970, Progress of
Theoretical Physics, 44, 291

\bibitem[Jaranowski et al.(1998)]{Jaranow98} Jaranowski, P.,
Kr{\'o}lak, A., \& Schutz, B.~F.\ 1998, \prd, 58, 063001

\bibitem[Jones
\& Andersson(2002)]{Jones02} Jones, D.~I., \& Andersson, N.\ 2002, \mnras, 331, 203

\bibitem[Katz et al.(1994)]{Katz94} Katz, J.~I., Toole, H.~A.,
\& Unruh, S.~H.\ 1994, \apj, 437, 727 

\bibitem[Kristian et al.(1989)]{Kristian89} Kristian, J.,
Pennypacker, C.~R., Morris, D.~E., et al.\ 1989, \nat, 338, 234

\bibitem[Krivoruchenko
\& Martem'ianov(1991)]{Krivoru91} Krivoruchenko, M.~I., \& Martem'ianov, B.~V.\ 1991, \apj, 378, 628

\bibitem[Lattimer
\& Prakash(2007)]{Lattimer07} Lattimer, J.~M., \& Prakash, M.\ 2007, \physrep, 442, 109

\bibitem[Lattimer et al.(1994)]{Lattimer94} Lattimer, J.~M., van
Riper, K.~A., Prakash, M., \& Prakash, M.\ 1994, \apj, 425, 802

\bibitem[Lindblom
\& Mendell(2000)]{Lindblom00} Lindblom, L., \& Mendell, G.\ 2000, \prd, 61, 104003

\bibitem[Madsen(1999)]{Madsen99} Madsen, J.\ 1999, Hadrons in
Dense Matter and Hadrosynthesis, 516, 162

\bibitem[Madsen(1998)]{Madsen98} Madsen, J.\ 1998, Physical
Review Letters, 81, 3311

\bibitem[Messenger
\& Read(2012)]{Messen12} Messenger, C., \& Read, J.\ 2012, Physical Review Letters, 108, 091101

\bibitem[Moraes
\& Miranda(2014)]{Moraes14} Moraes, P.~H.~R.~S., \& Miranda, O.~D.\ 2014, \mnras, 445, L11

\bibitem[Nakamura et al.(1997)]{Nakamura97} Nakamura, T., Sasaki,
M., Tanaka, T., \& Thorne, K.~S.\ 1997, \apjl, 487, L139 

\bibitem[Nissanke et al.(2010)]{Nissanke10} Nissanke, S., Holz,
D.~E., Hughes, S.~A., Dalal, N., \& Sievers, J.~L.\ 2010, \apj, 725, 496

\bibitem[Page
\& Applegate(1992)]{Page92} Page, D., \& Applegate, J.~H.\ 1992, \apjl, 394, L17

\bibitem[Panei et
al.(2000)]{Panei00} Panei, J.~A., Althaus, L.~G., \& Benvenuto, O.~G.\ 2000, \aap, 353, 970

\bibitem[Peters
\& Mathews(1963)]{Peters63} Peters, P.~C., \& Mathews, J.\ 1963, Physical Review, 131, 435

\bibitem[Pizzochero(1991)]{Pizzoch91} Pizzochero, P.~M.\ 1991,
Physical Review Letters, 66, 2425

\bibitem[Postnov
\& Yungelson(2014)]{Postnov14} Postnov, K.~A., \& Yungelson, L.~R.\ 2014, Living Reviews in Relativity, 17, 3

\bibitem[Press
\& Thorne(1972)]{Press72} Press, W.~H., \& Thorne, K.~S.\ 1972, \araa, 10, 335

\bibitem[Punturo et al.(2010)]{Punturo10} Punturo, M., Abernathy,
M., Acernese, F., et al.\ 2010, Classical and Quantum Gravity, 27, 084007

\bibitem[Schutz(1986)]{Schutz86} Schutz, B.~F.\ 1986, \nat, 323,
310

\bibitem[Timmes et al.(1996)]{Timmes96} Timmes, F.~X., Woosley,
S.~E., \& Weaver, T.~A.\ 1996, \apj, 457, 834

\bibitem[Vartanyan et al.(2014)]{Vartanyan14} Vartanyan, Y.~L.,
Hajyan, G.~S., Grigoryan, A.~K.,
\& Sarkisyan, T.~R.\ 2014, Journal of Physics Conference Series, 496, 012009

\bibitem[Witten(1984)]{Witten84} Witten, E.\ 1984, \prd, 30, 272

\bibitem[Wolszczan
\& Frail(1992)]{Wolszczan92} Wolszczan, A., \& Frail, D.~A.\ 1992, \nat, 355, 145

\bibitem[Xu(2006)]{Xu06} Xu, R.~X.\ 2006, Astroparticle
Physics, 25, 212

\bibitem[Xu
\& Wu(2003)]{Xu03} Xu, R.-X., \& Wu, F.\ 2003, Chinese Physics Letters, 20, 806 
\end{thebibliography}
\end{document}